\documentstyle[twocolumn,prl,aps,epsfig,amsfonts]{revtex}

\begin{document}
\draft
\preprint{}

\newcommand{\1}{{\bf \scriptstyle 1}\!\!{1}}
\newcommand{\I}{{\rm i}}
\newcommand{\p}{\partial}
\newcommand{\D}{^{\dagger}}
\newcommand{\bx}{{\bf x}}
\newcommand{\bk}{{\bf k}}
\newcommand{\bv}{{\bf v}}
\newcommand{\bp}{{\bf p}}
\newcommand{\bu}{{\bf u}}
\newcommand{\bA}{{\bf A}}
\newcommand{\bB}{{\bf B}}
\newcommand{\bF}{{\bf F}}
\newcommand{\bI}{{\bf I}}
\newcommand{\bK}{{\bf K}}
\newcommand{\bL}{{\bf L}}
\newcommand{\bP}{{\bf P}}
\newcommand{\bQ}{{\bf Q}}
\newcommand{\bS}{{\bf S}}
\newcommand{\bH}{{\bf H}}
\newcommand{\balpha}{\mbox{\boldmath $\alpha$}}
\newcommand{\bsigma}{\mbox{\boldmath $\sigma$}}
\newcommand{\bSigma}{\mbox{\boldmath $\Sigma$}}
\newcommand{\bOmega}{\mbox{\boldmath $\Omega$}}
\newcommand{\bpi}{\mbox{\boldmath $\pi$}}
\newcommand{\bphi}{\mbox{\boldmath $\phi$}}
\newcommand{\bnabla}{\mbox{\boldmath $\nabla$}}
\newcommand{\bmu}{\mbox{\boldmath $\mu$}}
\newcommand{\bepsilon}{\mbox{\boldmath $\epsilon$}}

\newcommand{\iLambda}{{\it \Lambda}}
\newcommand{\cA}{{\cal A}}
\newcommand{\cD}{{\cal D}}
\newcommand{\cF}{{\cal F}}
\newcommand{\cL}{{\cal L}}
\newcommand{\cH}{{\cal H}}
\newcommand{\cI}{{\cal I}}
\newcommand{\cM}{{\cal M}}
\newcommand{\cO}{{\cal O}}
\newcommand{\cR}{{\cal R}}
\newcommand{\cU}{{\cal U}}
\newcommand{\cT}{{\cal T}}

\newcommand{\be}{\begin{equation}}
\newcommand{\ee}{\end{equation}}
\newcommand{\bea}{\begin{eqnarray}}
\newcommand{\eea}{\end{eqnarray}}
\newcommand{\beqa}{\begin{eqnarray*}}
\newcommand{\eeqa}{\end{eqnarray*}}
\newcommand{\nn}{\nonumber}
\newcommand{\DD}{\displaystyle}

\newcommand{\ba}{\left[\begin{array}{c}}
\newcommand{\baa}{\left[\begin{array}{cc}}
\newcommand{\baaa}{\left[\begin{array}{ccc}}
\newcommand{\baaaa}{\left[\begin{array}{cccc}}
\newcommand{\ea}{\end{array}\right]}

\twocolumn[
\hsize\textwidth\columnwidth\hsize\csname
@twocolumnfalse\endcsname

\title{Quantum information processing with large nuclear spins in GaAs
semiconductors}

\author{Michael N.~Leuenberger %\cite{email1} 
and Daniel Loss%\cite{email2}
}
\address{Department of Physics and Astronomy, University of Basel \\
Klingelbergstrasse 82, 4056 Basel, Switzerland}
\author{Martino Poggio %\cite{email3} 
and David D.~Awschalom%\cite{email4}
}
\address{Department of Physics, University of California, Santa Barbara,
CA 93106-9530, USA}

\date{\today}
\maketitle

\begin{abstract}
We propose an implementation for quantum information processing based on
coherent manipulations of nuclear spins $I=3/2$ in GaAs semiconductors.
We describe theoretically an NMR method which
involves multiphoton transitions and which exploits
the non-equidistance of nuclear spin levels due to quadrupolar splittings.
Starting from known spin anisotropies we derive effective
Hamiltonians in a generalized rotating frame, valid for arbitrary $I$,
which allow us to describe the non-perturbative time evolution of spin states
generated by magnetic rf fields.
We identify an experimentally accessible regime where
multiphoton Rabi oscillations are
observable. In the nonlinear regime, we find Berry phase
interference effects.
\end{abstract}

]
\narrowtext

Recent advances in spintronics\cite{Wolf}  have shown that
the coherent control of electron and nuclear spins in
semiconductors  is experimentally feasible,
enabling in particular an all-optical NMR in GaAs, based on the hyperfine interaction
between electrons and nuclei\cite{Kikkawa2000,Salis}.
Such a control of nuclear spins 
can also be achieved via electrical gates
as recently demonstrated for GaAs heterostructures in the  quantum
Hall regime\cite{Smet},
or even via conventional NMR techniques directly accessing the
nuclei\cite{Abragam}.
In the present work, we will show  that  such advances in coherent spin control
have opened up
the possibility to  manipulate the nuclear
spins $I$ for the purpose of quantum information processing, thereby presenting a
scheme that
is based on ensembles of large spins $I>1/2$ instead
of qubits.
Nuclear spins are ideal candidates for this purpose due to their long decoherence
times.

An implementation of the Grover algorithm\cite{Grover} has
recently been proposed for molecular magnets\cite{Leuenberger&Loss}, based on a
perturbative approach to the unitary Grover operations
which encode and decode the information stored in the phases of {\em small} amplitudes $a_m$\cite{Ahn}.
An alternative version of Grover's algorithm was presented in Refs.~\cite{Farhi,Grover&Sengupta}
that is described by a Hamiltonian 
that lets a completely delocalized state
of the form $\left|\psi\right>=\sum_{m=-I}^Ia_m\left|m\right>$, in some  basis
states $\left|m\right>$ with equal occupation probabilities $|a_m|^2$, evolve into
a wanted localized state $\left|M\right>$,
where $\left|\psi\right>$ and $\left|M\right>$ are degenerate and have a finite overlap.
The information is encoded in the energies of $\left|m\right>$.
In order to produce $\left|\psi\right>$, we propose here a novel NMR scheme that
allows us to generate any desired
distribution of amplitudes $a_m$, being not restricted to small values.
For this we specifically  exploit the properties
of GaAs nuclei where quadrupolar spin splitting results in spin anisotropies
and thus in non-equidistant energy levels--
being a necessary condition for our scheme. The theoretical problem then is to
find one magnetic rf pulse---inducing a unitary time evolution of 
the spins---that produces the desired spin state $\left|\psi\right>$
and a second rf pulse that lets $\left|\psi\right>$ evolve into $\left|M\right>$, 
given certain spin anisotropies and adjustable magnetic
fields (see below). In a non-perturbative approach,  we  find  an analytic
solution to this problem, valid for arbitrary spin $I$.  
For the special case of GaAs with $I=3/2$, we have
confirmed our analytical results by exact numerics.
In contrast to previous
work\cite{Leuenberger&Loss} our method also holds for vanishing
detuning energies, which turns out to be essential to perform 
non-perturbative unitary operations, i.e. quantum computations (QCs).  
Once the control over $2I$ magnetic fields is established, the scheme
proposed here allows for QC and quantum storage with a {\em single}
pulse, provided that there is sufficient signal
amplification due to the spin ensemble\cite{footnote1}.

\begin{figure}[htb]
  \begin{center}
    \leavevmode
\epsfxsize=8cm
\epsffile{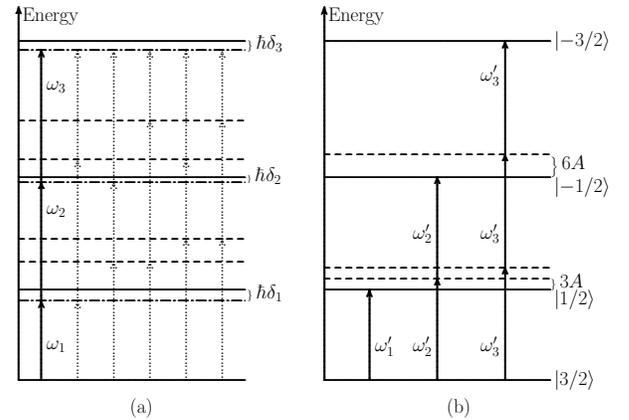}
  \end{center}
\caption{Multiphoton transition schemes for the coherent population of
the $I_z$ eigenstates $\left|m\right>$ of a nuclear spin $I=3/2$. 
(a) Quantum computation (QC) scheme:
The frequencies $\omega_k$ of the  fields
$H_k$ are red (- $\cdot$) and blue
(- -) detuned. Diagrams containing blue detunings are negligible for
large quadrupolar splitting, i.e. $A\gg\hbar\delta_k\ge 0$. 
(b) Rabi oscillation (RO) scheme: The magnetic
fields $H_k'\cos(\omega_k't+\Phi_k')$, 
$k=1,2,3,$ give rise to $k$-photon RO.}
\label{nuclear_spin_spectrum}
\end{figure}

%in a single-spin Hilbert space of dimension $2I+1$.
As a first step towards this goal, it will be useful to generate and monitor
multiphoton Rabi oscillations, as we describe
in detail below.
Finally, we show that oscillating
quadratic transverse spin terms, which can be generated by optical pulses in
GaAs\cite{Salis,Brun},
give rise to Berry phase oscillations\cite{LLBerry} in the transition
probabilities.

In the following we mainly focus on a nuclear spin of length $I=3/2$, as appropriate for GaAs,
but indicate its  generalization to arbitray $I$.
Our nuclear spin system is  described by the Hamiltonian
$\cH_0=\cH_{\rm Z}+\cH_{\rm Q}$, consisting of the nuclear Zeeman energy
$\cH_{\rm Z}=-g_N\mu_NH_zI_z$, 
$g_N=1.3$\cite{Kikkawa2000}, and the quadrupolar splitting\cite{Abragam} 
$\cH_{\rm Q}=A[3I_z^2-I(I+1)]$. The quadrupolar constant is 
$A=7\times 10^{-7}$ K for $^{69}$Ga, $A=3\times 10^{-7}$ K for $^{71}$Ga, 
and $A=2\times 10^{-6}$ K for $^{75}$As nuclei\cite{Salis}.
For the purpose of QC we need to achieve complete control over unitary state evolutions, i.e.
control over amplitudes $a_m$ to form a desired superposition
$\left|\psi\right>=\sum_{m=-I}^Ia_m\left|m\right>$ of the nuclear basis states 
$\left|m\right>$ (eigenstates of $\cH_0$). 
Our goal is now to show that such a control over $a_m$
is indeed feasible under experimentally attainable conditions. 

\begin{figure}[htb]
  \begin{center}
    \leavevmode
\epsfxsize=7cm
\epsffile{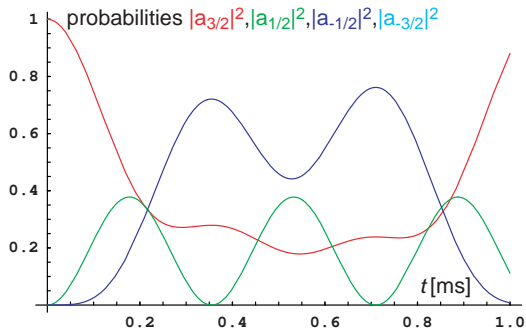}
  \end{center}
\caption{Preparation of 
$\left|s\right>=(1/\protect\sqrt{3})\sum_{m=-1/2}^{3/2}\left|m\right>$
by means of Eq.~(2)
in the QC scheme, which takes about 0.2 ms for $H_1=H_2=1$ G,
$H_3=0$, $\delta_1=6083$ s$^{-1}$, and $\delta_2=0$.
The duration of the QC operation is 
$<1/2\nu_{\rm Rabi}^{(2)}$. 
The analytical result is confirmed by numerics.
}
\label{QC}
\end{figure}

We start from a configuration where mainly the ground state
$\left|3/2\right>$ is populated, see Fig.~\ref{nuclear_spin_spectrum}.
 This can be achieved by the
Overhauser effect\cite{Overhauser}. 
The next goal is to coherently populate all or a part of the
excited states $\left|m\right>$, $m\ne 3/2$, by means of $\Delta
m=1,2,3$-photon transitions. 
Fig.~\ref{nuclear_spin_spectrum} shows the two transition
schemes, QC and RO, which will turn out to be appropriate for 
quantum computation (QC) and multiphoton Rabi oscillations (RO), resp.  
In the QC scheme the frequencies $\omega_k$ of the external
transverse magnetic fields,
$H_{x,k}(t)=\tilde{H}_k(t)\cos(\omega_kt+\Phi_k)$, $k=1,2,3$, 
are blue ($\delta_k<0$) and red ($\delta_k>0$)
detuned with respect to the energy level separations
$\hbar\omega_{m,m'}$. In the RO scheme, the transverse
fields $H_{x,k}'(t)=\tilde{H}_k'(t)\cos(\omega_k't+\Phi_k')$, $k=1,2,3$,
oscillate at frequencies 
$\omega_{\Delta m}'=\omega_{3/2-\Delta m,3/2}/\Delta m$, 
which are blue detuned by 
$3A$ ($6A$) for the two(three)-photon transition. 
For  GaAs, $\omega_k,\omega_k'\sim 10$ MHz with
$\delta_k\sim 1$ kHz,
and a longitudinal magnetic field $H_z\sim 1$ T is appropriate.
It is desirable to make $H_z$ sufficiently large to accommodate many spin precessions
before the spins dephase.
We note that in contrast to the fields
$H_{x,k}(t)$, the fields 
$H_{x,k}'(t)$ lead to transitions governed by non-commuting operators,
with the important consequence that the RO scheme suffers from strong interferences 
between the transitions if two or more fields $H_{x,k}'(t)$ are nonzero,
leading to a quick loss of amplitude control.
Indeed, the RO scheme  allows control of $a_m\,'s$ only for times 
$t\ll 2\hbar(V_k'+V_{k'}')/V_k'V_{k'}'$, which we estimate from the Baker-Campbell-Hausdorff formula 
and which we confirmed by exact numerical calculations. Here,
$V_k'=2[(g_N\mu_NH_k')^kp_{3/2-k,3/2}]/\prod_{j=1}^{k-1}\hbar\omega_{\frac{3}{2}-j,\frac{3}{2}}$
(see below).  Although the RO scheme is only suited for QCs using
perturbative approaches, it has its usefulness for testing the coherence of the spin system (see below).

\begin{figure}[htb]
  \begin{center}
    \leavevmode
\epsfxsize=7cm
\epsffile{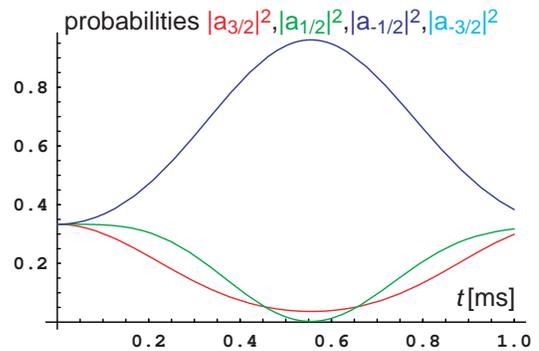}
  \end{center}
\caption{Grover algorithm calculated by means of 
Eq.~(2) 
in the QC scheme (numerically confirmed), where  
$\left|s\right>=(1/\protect\sqrt{3})\sum_{m=-1/2}^{3/2}\left|m\right>$ 
is concentrated mainly
into $\left|-1/2\right>$ after 0.55 ms for $H_2=\protect\hbar\delta_2/2g_N\mu_N=1$ G, 
$h_1=h_2$, $h_3=0$, $\delta_1=0$. 
The duration of the QC is $<1/2 \nu_{\rm Rabi}^{(2)}$.
}
\label{Grover}
\end{figure}

Now we proceed with demonstrating the existence of the desired
spin transitions in the QC scheme. For this we evaluate the transition amplitudes for the
diagrams of Fig. 1 (a) in high-order
perturbation theory which allows us then to obtain an appropriate non-perturbative
Hamiltonian (see below).
The three transverse fields $H_{x,k}(t)$ complete the Hamiltonian
$\cH=\cH_0+V(t)$, where
$
V(t)=\sum_{k=1}^3 g_N\mu_N\tilde{H}_k(t)\cos(\omega_kt+\Phi_k)I_x,
$
with $I_x=(I_++I_-)/2$, and phases $\Phi_k$ (see below).  
Then we expand the $S$-operator $S=\sum_{j=0}^\infty S^{(j)}$ in powers
of  
$V(t)$. 
We use rectangular pulse shapes of duration $T$ for all fields, i.e.
$\tilde{H}_k(t)=H_k$ for $-T/2<t<T/2$, and $0$ otherwise.
Then we obtain
\bea
\tilde{S}_{-\frac{3}{2},\frac{3}{2}}^{(3)} & = & 
\prod_{k=1}^3H_ke^{-i\Phi_k}
\left[\frac{1}{\delta_1\delta_2}-\frac{1}{\delta_1(\frac{6A}{\hbar}-\delta_1+\delta_2)}\right.
\nn\\
& &
-\left.\frac{1}{\frac{6A}{\hbar}+\delta_1-\delta_2}\left(
\frac{1}{\delta_2}
-\frac{1}{\frac{12A}{\hbar}+\delta_1}\right)\right.
\nn\\
& &
+\left.\frac{1}{\frac{12A}{\hbar}+\delta_2}\left(
\frac{1}{\frac{6A}{\hbar}-\delta_1+\delta_2}
+\frac{1}{\frac{12A}{\hbar}+\delta_1}\right)\right]
\label{S3}
\eea
for $\delta_3=0$,
$
\tilde{S}_{-\frac{1}{2},\frac{3}{2}}^{(2)}=
\prod_{k=1}^2H_ke^{-i\Phi_k}
\left(-\frac{1}{\delta_1}+\frac{1}{\frac{6A}{\hbar}+\delta_1}\right)
$
for $\delta_2=0$ and $H_3=0$, and 
$
\tilde{S}_{\frac{1}{2},\frac{3}{2}}^{(1)}=H_1e^{-i\Phi_1}
$
for $\delta_1=0$ and $H_2=H_3=0$,
where
$S_{\frac{3}{2}-j,\frac{3}{2}}^{(j)}=\frac{2\pi}{i}
\left(\frac{g_N\mu_N}{4\hbar}\right)^j\tilde{S}_{\frac{3}{2}-j,\frac{3}{2}}^{(j)}
p_{\frac{3}{2}-j,\frac{3}{2}}
\delta^{(T)}(\omega_{\frac{3}{2}-j,\frac{3}{2}}-\sum_{k=1}^j\omega_k)$,
$p_{m,m'}=\prod_{k=m}^{m'}\left<k\left|I_-\right|k+1\right>$, 
and
$\delta^{(T)}(\omega)=\frac{1}{2\pi}\int_{-T/2}^{+T/2}e^{i\omega
t}dt=\frac{\sin(\omega T/2)}{\pi\omega}$ is the delta-function of width
$1/T$. The energy is conserved for $\omega T\gg 1$. Also, the duration
$T$ of the rf pulses must not exceed the dephasing time $\tau_{\phi}$ of 
the spin states.
%, and the pulses must be switched on and off adiabatically.
Interestingly, $\lim_{A\rightarrow 0} S_{-3/2,3/2}^{(3)}=$
$\lim_{A\rightarrow 0} S_{-1/2,3/2}^{(2)}=0$, i.e. destructive
interference is maximal. However, if $A\gg\hbar|\delta_k|$, $k=1,2,3$,
destructive interference is negligible.
%i.e. only the first terms in
%Eqs.~(\ref{S3}) and (\ref{S2}) give a significant contribution to 
%$S_{-\frac{3}{2},\frac{3}{2}}^{(3)}$ and 
%$S_{-\frac{1}{2},\frac{3}{2}}^{(2)}$, resp. 
%If the energy of the final state $\left|m\right>$ is inhomogeneously
%distributed, we must %integrate $|S_{m,I}^{(j)}|^2$ over the Gaussian
%energy distribution 
%$\rho(\omega)=(1/\sqrt{2\pi}\Delta\omega)
%\exp\{-(\omega-\omega_{m,I})^2/(2\Delta\omega^2)\}$, i.e. the total
%transition probability
%$P_{\rm tot}=\sum_{m',\pm}|S_{m',\frac{3}{2}}^{(j)}|^2$ is given by
%$
%P_{\rm tot} = 
%2\pi\left(\frac{g_N\mu_N}{4\hbar}\right)^{2j} 
%\left|\tilde{S}_{\frac{3}{2}-j,\frac{3}{2}}^{(j)}p_{\frac{3}{2}-j,\frac{3}{2}}\right|^2
%\rho(\omega)T,
%$
%where
%$\rho(\omega_{\frac{3}{2}-j,\frac{3}{2}})=1/\sqrt{2\pi}\Delta\omega$.

\begin{figure}[htb]
  \begin{center}
    \leavevmode
\epsfxsize=7cm
\epsffile{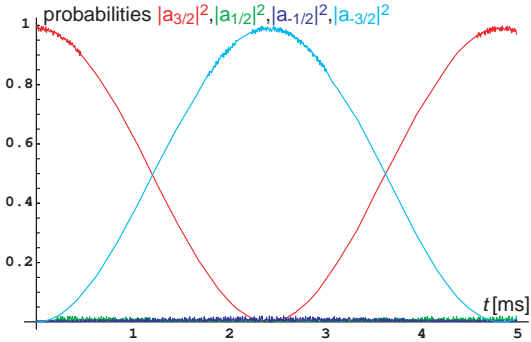}
  \end{center}
\caption{Numerical solution for the three-photon ROs of $^{75}$As nuclei 
between $\left|3/2\right>$ and
$\left|-3/2\right>$ driven by $H_3'=20$ G with the RO scheme (b)
in cw mode. $H_1'=H_2'=0$.}
\label{Rabi_3photon}
\end{figure}
 
Now we are in the position to extract an effective Hamiltonian, 
which governs the desired unitary evolutions\cite{Farhi,Grover&Sengupta}. 
For
the QC scheme we use the Hamiltonian $\cH$.  
After applying the rotating wave approximation\cite{Abragam} we
keep only the most left
diagram of Fig. 1(a), which gives the dominant contribution to the
transition amplitudes for $\hbar|\delta_k|\ll |A|$. 
This is a direct consequence of the non-equidistance of the energy levels
$\left|m\right>$ due to the quadrupolar splitting. It is now possible to eliminate the 
time-dependence of $\cH$ by a unitary operation $U(t)$, the matrix elements
of which can be determined by solving $2I$ linear equations.
This is a transformation to a generalized rotating frame.
Then, for a spin  $I$ we obtain an effective time-independent Hamiltonian 
\be
\cH_{\rm rot}^{(2I)}=\left[\begin{array}{ccccc} 
0 & h_1 & 0 & \cdots & 0 \\
h_1 & \hbar\delta_1 & h_2 & \ddots & \vdots \\
0 & h_2 & \hbar\delta_2 & \ddots & 0  \\
\vdots & \ddots & \ddots & \ddots & h_{2I} \\ 
0 & \cdots & 0 & h_{2I} & \hbar\delta_{2I} \ea,
\label{generalrot}
\ee
where $h_k=g_N\mu_NH_k\sqrt{k(2I+1-k)}/2$ ($k=1,\ldots,2I$).
Focusing on $I=3/2$, we obtain e.g. for $H_3=0$ approximately 
$\cH^{(2)}=\cH_0+h_1e^{i(\omega_1t+\Phi_1)}\left|3/2\right>\left<1/2\right|$
$+h_2e^{i(\omega_2t+\Phi_2)}\left|1/2\right>\left<-1/2\right|$
$+{\rm h.c.}$. Applying 
$U(t)=e^{-i[(\omega_1+\omega_2)t+(\Phi_1+\Phi_2)]/2}\left|3/2\right>\left<3/2\right|
+e^{i[(\omega_1-\omega_2)t+(\Phi_1-\Phi_2)]/2}
\left|1/2\right>\left<1/2\right|
+e^{i[(\omega_1+\omega_2)t+(\Phi_1+\Phi_2)]/2}
\left|-1/2\right>\left<-1/2\right|$
yields $\cH_{\rm rot}^{(2)}$.
%\be
%\cH_{\rm rot}^{(2)}=\baaa 0 & \frac{\sqrt{3}}{2}g_N\mu_N H_1 & 0 \\
%\frac{\sqrt{3}}{2}g_N\mu_N H_1 & \hbar\delta_1 & g_N\mu_N H_2 \\
%0 & g_N\mu_N H_2 & \hbar\delta_2 \ea.
%\label{rotating_frame_approx}
%\ee
Note that the Hamiltonian in Eq.~(\ref{generalrot}) remains valid
even in the limit $\delta_k\rightarrow 0$, where perturbation
expansions such as in Eq.~(\ref{S3}) break down.
However, we must require that $|g_N\mu_N H_k|\ll |A|$, which means that
the larger $|A|$, the faster the QCs. 
Propagators of the
form $U^\dagger(t)e^{-i\cH_{\rm rot}^{(2I)}t/\hbar}$ have $2I$
phases $\Phi_k$ and $2I$ detunings
$\hbar\delta_k$, which determine the $2I$
phases and the $2I$ moduli of $a_m$\cite{freedom}.  
%To start and end a QC, $H_k$ are switched on and off
%adiabatically. For further control the phases
%$\Phi_k(t)$ can be changed adiabatically ($\dot{\Phi}_k\ll\omega_k$).

For Grover's algorithm\cite{Farhi,Grover&Sengupta} we must first
produce $\left|s\right>=(1/\sqrt{n})\sum_m\left|m\right>$
(see Fig.~\ref{QC}),
$n$ being the number of basis states involved in the search.
Then we make use of
the degeneracy between $\left|s\right>$
and $\left|M\right>$, which yields the resonance condition
$h_k=\hbar\delta_{I-M}/2\ne 0\;\forall k$, if $\delta_k=0\;\forall k\ne I-M$.
In contrast to \cite{Farhi,Grover&Sengupta},
$\cH_{\rm rot}^{(2I)}$ has only nearest-neighbor coupling,
which results in a decreasing amplification of $\left|M\right>$
with increasing $I$ or $|M|$. However, even for the largest nuclear spin
$I=9/2$, we find that the resolution for identifying $\left|M\right>$
is still sufficient ($\gtrsim 10$\%).
Fig.~\ref{Grover} shows the example where $\left|M=-1/2\right>$ is found
out of the three states $\left|m\right>$, $m=3/2,1/2,-1/2$, for $I=3/2$.

\begin{figure}[htb]
  \begin{center}
    \leavevmode
\epsfxsize=7cm
\epsffile{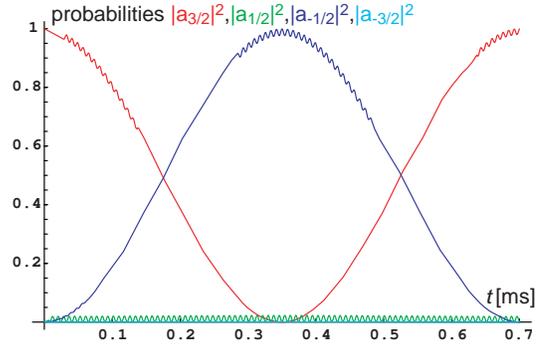}
  \end{center}
\caption{Numerical solution for the two-photon ROs
of $^{75}$As nuclei between $\left|3/2\right>$ and $\left|-1/2\right>$,
driven by $H_2'=10$ G according to the RO scheme in cw mode, and $H_1'=H_3'=0$. }
\label{Rabi_2photon}
\end{figure}

As a first test for the proposed schemes, it would be useful to measure
generalized ROs involving multiphoton absorptions. They can be thought of as nutation of
the large spin $I$ between spin states $\left|m\right>$.
First, we consider the QC scheme.
For the two-photon RO, with frequency $\nu_{\rm Rabi}^{(2)}$, to become observable, 
we need $|g_N\mu_N
H_k|\ll
\hbar\delta_1$, $k=1,2$, so that the one-photon transitions
are completely suppressed. To obtain $\nu_{\rm Rabi}^{(2)}$, it is
useful to think of  (\ref{generalrot})
as describing the  dynamics of a (fictitious) particle in a triple well
with nearest-neighbor tunnel coupling $h_k$. The independent control of the
tunnel couplings $h_k$ and the biases $\hbar\delta_k$ between the wells is ensured 
by a large value of $A$.
Then, the energy (``tunnel") splitting\cite{LLMn12} between
$\left|3/2\right>$ and $\left|-1/2\right>$ reads for $\delta_2=0$
\be
\Delta_{\rm Rabi}^{(2)} = {\sqrt{3}(g_N\mu_N)^2H_1H_2}/{\delta_1},
\label{Rabi_splitting}
\ee
which gives 
$\nu_{\rm Rabi}^{(2)}=\Delta_{\rm Rabi}^{(2)}/2\pi\hbar$.
%Even if $|H_1|,|H_2|\gtrsim \hbar\delta_1/g_N\mu_N$,
%Eq.~(\ref{Rabi_splitting}) gives the %envelope frequency of the
%two-photon Rabi oscillation, whose fast oscillation frequency can be
%interpreted as the renormalization of the levels. 
In order to obtain large Rabi frequencies $\nu_{\rm Rabi}^{(2)}$, the
external  fields $H_1,H_2$ and the detuning $\hbar\delta_1$ are
to be maximized under the conditions $|H_1|,|H_2|\ll
\hbar|\delta_1|/g_N\mu_N\ll |A|/g_N\mu_N$\cite{condition}, i.e. 
the larger $|A|$ the larger $\nu_{\rm
Rabi}^{(2)}$ can be 
achieved. We note that $|A|$ could e.g. be enhanced by optical laser pumping\cite{Salis} or by
modulated electric field gradients\cite{Brun}.

Next we turn to the RO scheme. Here  it is sufficient to apply only one single
field $H_{x,k}'(t)$ in order to see the multiphoton ROs 
shown in Figs.~\ref{Rabi_3photon}, \ref{Rabi_2photon}.
%and the corresponding Rabi phase oscillation shown in Fig.~\ref{Rabi_phase}.
We now also allow for  oscillating quadratic transverse anisotropies which
can be externally generated
by modulating 
the electric field gradient felt by the nuclei\cite{Salis,Brun}. For
this we adopt the most general Hamiltonian\cite{Abragam}
\bea
\cH' & = &
A[3I_z^2-I(I+1)]-g_N\mu_NH_zI_z+e^{i\omega_k'tI_z}\left[-h_k'I_x\right.
\nn\\
& &
+\left.B\left(I_xI_z+I_zI_x\right)+C\left(I_x^2-I_y^2\right)\right]e^{-i\omega_k'tI_z},
\eea
where $h_k'=g_N\mu_NH_k'$ ($k=1,2$ or $3$).
Next we transform $\cH'$
to the rotating frame, which yields $\cH{'}_{\rm rot}=A[3I_z^2-I(I+1)]
-(g_N\mu_NH_zI_z-\hbar\omega_k')I_z+B\left(I_xI_z+I_zI_x\right)
+C\left(I_x^2-I_y^2\right)+h_k'I_x$.
Then the time evolution takes the simple form 
$\left|\psi(t)\right>=e^{i\omega_k'tI_z}e^{-i\cH{'}_{\rm
rot}t/\hbar}\left|I\right>$.
Although the transverse quadratic term $C$ is not in resonance with any
transition energy, it leads to a time-independent transverse quadratic
anisotropy in the rotating frame. 
For the 3-photon transition in the RO scheme  we obtain the
following Hamiltonian in the rotating frame
\be
\cH{'}_{\rm rot}^{(3)}=\baaaa 3A & \frac{\sqrt{3}}{2}h_3' & \sqrt{3}C &
0 \\ 
\frac{\sqrt{3}}{2}h_3' & -3A & h_3' & \sqrt{3}C \\
\sqrt{3}C & h_3' & -3A & \frac{\sqrt{3}}{2}h_3' \\ 
0 & \sqrt{3}C & \frac{\sqrt{3}}{2}h_3' & 3A \ea,
\label{rotating_frame}
\ee
where we have neglected the $B$ term since we choose $B\ll h_3'$.
Inserting a typical value $C=-10^{-10}$ K\cite{Salis}, we obtain
oscillations of the splitting 
$\Delta{'}_{\rm Rabi}^{(3)}$ between $\left|3/2\right>$ and
$\left|-3/2\right>$ as a function of $H_3'$, see
Fig.~\ref{Berryphase_oscillations}. These oscillations are due to the
Berry phase in a biaxial spin system as shown in \cite{LLBerry}. Note
that $C$ must be negative for the Berry phase interference to
occur\cite{LLBerry}. Also, the Berry phase interference is present only
for $\Delta m$-photon transitions with $\Delta m\ge 2$.
In Figs.~\ref{Rabi_3photon} and \ref{Rabi_2photon} the population
probabilities 
$|a_m(t)|^2$ are shown for $C=0$. The corresponding normalized
magnetization reads
$M(t)=\sum_m m|a_m|^2$.

%For completeness we give also the Hamiltonian for the 2-photon process
%in the rotating frame
%\be
%\cH_{\rm rot}^{(2)}=\baaa -3A/2 & \frac{\sqrt{3}}{2}g_N\mu_N H_2' & 0
%\\ 
%\frac{\sqrt{3}}{2}g_N\mu_N H_2' & -9A/2 & g_N\mu_N H_2' \\
%0 & g_N\mu_N H_2' & -3A/2 \ea.
%\label{rotating_frame_2photon}
%\ee

%Alternatively the $B$ and $C$ terms could be used to control the phases
%and populations of the %$\left|m\right>$ states in $\left|\psi\right>$. 
%This approach has the advantage that for the $\Delta m=\pm 2$
%transition there is no leakage %to the $\Delta m=\pm 1$ transition.
%However, the effective fields that can currently be %achieved are of the
%order of $10^{-6}$ T\cite{Kikkawa2000}.

In order to build parallel single-spin quantum computers or high-density
memory devices, one could apply an inhomogeneous magnetic field to the
GaAs sample. Then the nuclear Larmor frequencies are spread over a wide
range, which could be divided into small frequency intervals that are
individually accessible by NMR sources.

In conclusion we have shown that via multiphoton transitions a
controlled superposition of  spin states can be achieved by
 appropriate field pulses. 
%This provides the
%possibility to perform any QC on an
%ensemble of large spins with sufficient signal amplification. 
%Due to their long dephasing
%time nuclear spins of GaAs crystals are promising candidates. 
Our
results can be extended to arbitrary spin $I$ and 
to any single-particle quantum system with non-equidistant energy
levels.

\begin{figure}[htb]
  \begin{center}
    \leavevmode
\epsfxsize=7cm
\epsffile{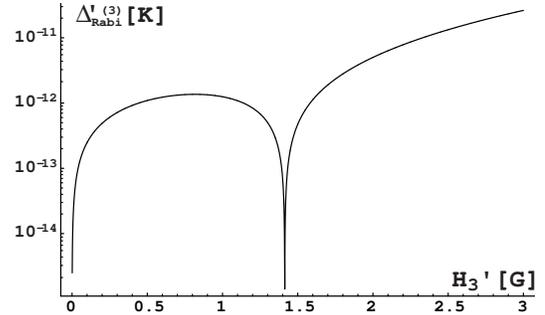}
  \end{center}
\caption{Berry phase oscillation. The three-photon transition
probability vanishes where $\Delta{'}_{\rm Rabi}^{(3)}$ is zero.}
\label{Berryphase_oscillations}
\end{figure}

{\it Acknowledgement}. We thank A.~J.~Leggett and G.~Salis for useful
discussions. We acknowledge the Swiss NSF, NCCR Nanoscience, Molnanomag, DARPA, and ARO
for financial support. This research was supported in part by the
US NSF under Grant No. PHY99-07949.


\begin{references}
\vspace{-0.3cm}
%\bibitem[*]{email1} Electronic address: Michael.Leuenberger@unibas.ch
%\bibitem[\dagger]{email2} Electronic address: Daniel.Loss@unibas.ch
%\bibitem[**]{email3} Electronic address: poggio@iquest.ucsb.edu
%\bibitem[\ddagger]{email4} Electronic address: awsch@physics.ucsb.edu
\bibitem{Wolf} %G.~A.~Prinz, Phys. Today {\bf 45}(4), 58 (1995); 
%G.~A.~Prinz, Science {\bf 282}, 1660 (1998); 
S.~A.~Wolf {\it et al.}, Science {\bf 294}, 1488 (2001).
%\bibitem{Kikkawa1999} J.~M.~Kikkawa, D.~D.~Awschalom, Nature {\bf 397},
%139 (1999). 
%D.~D.~Awschalom, J.~M.~Kikkawa, Phys. Today {\bf 52}(6), 33 (1999).
%\bibitem{Kikkawa1997} J.~M.~Kikkawa et al., 
%I.~P.~Smorchkova, N.~Samarth, and D.~D.~Awschalom, 
%Science {\bf 277}, 1284 (1997). 
%J.~M.~Kikkawa, D.~D.~Awschalom, Phys.~Rev.~Lett. {\bf 80}, 4313 (1998).
\bibitem{Kikkawa2000} J.~M.~Kikkawa, D.~D.~Awschalom, Science {\bf 287},
473 (2000).
\bibitem{Salis} G.~Salis
%, D.~T.~Fuchs, J.~M.~Kikkawa, D.~D.~Awschalom, Y.~Ohno, H.~Ohno, 
{\it et al.}, Phys.~Rev.~Lett. {\bf 86}, 2677 (2001); 
G.~Salis {\it et al.}, 
%D.~D.~Awschalom, Y.~Ohno, H.~Ohno, 
Phys.~Rev.~B {\bf 64}, 195304 (2001).
\bibitem{Smet} J.~H.~Smet {\it et al.}, Nature {\bf 415}, 281 (2002). 
\bibitem{Abragam} A.~Abragam, {\it The Principles of Nuclear Magnetism}
(Clarendon, 1961).
\bibitem{Grover} L.~K.~Grover, Phys.~Rev.~Lett.~{\bf 79}, 4709 (1997).
\bibitem{Leuenberger&Loss} M.~N.~Leuenberger, D.~Loss, Nature {\bf 410},
789 (2001).
\bibitem{Ahn} J.~Ahn {\it et al.}, %T.~C.~Weinacht, P.~H.~Bucksbaum, 
Science {\bf 287}, 463 (2000).
\bibitem{Farhi} E.~Farhi, S.~Gutmann, Phys.~Rev.~A {\bf 57}, 2403 (1998).
\bibitem{Grover&Sengupta} L.~K.~Grover, A.~M.~Sengupta, Phys.~Rev.~A {\bf 65}, 032319 (2002).
\bibitem{footnote1} Like all implementations based on a
``unary" representation, the present scheme is not scalable.  
\bibitem{Brun} E.~Brun {\it et al.}, Phys.~Rev.~Lett. {\bf 8}, 365
(1962).
\bibitem{LLMn12} M.~N.~Leuenberger, D.~Loss, Phys.~Rev.~B {\bf 61}, 1286
(2000).
\bibitem{LLBerry} 
D.~Loss {\it et al.}, %D.~P.~DiVincenzo, G.~Grinstein, 
Phys. Rev. Lett. {\bf 69}, 3232 (1992); 
J.~von Delft, C.~L.~Henley, {\it ibid.}, 3236 (1992);
M.~N.~Leuenberger, D.~Loss, Phys.~Rev.~B {\bf 63}, 054414 (2001).
\bibitem{Overhauser} S.~E.~Barrett {\it et al.}, Phys.~Rev.~Lett. {\bf 72},
1368 (1994); J.~A.~Marohn {\it et al.}, Phys.~Rev.~Lett. {\bf 75}, 1364
(1995).
\bibitem{freedom} We subtracted two
degrees of freedom: the global phase and the normalization condition,
respectively.
\bibitem{condition} This condition must also be satisfied by the
numerical values in Ref.~\cite{Leuenberger&Loss}.
\end{references}
\end{document}